\documentclass{report} 
\usepackage{graphicx}

\renewcommand{\abstract}[1]{{ \footnotesize \noindent \textbf{Abstract} #1\\}}

\renewcommand{\author}[1]{\subsection*{#1}}
\newcommand{\address}[1]{\subsection*{\it#1}}

\begin{document}

\chapter*{The promise of interferometry in the mm range: ALMA\footnote{Contribution to XIII
Rencontres de Blois 2001, ed. L. M. Celnikier}}
\author{Alain Omont}


\address{Institut d'Astrophysique de Paris, 98bis boulevard  
Arago, F-75014 Paris}

\abstract{The Atacama Large Millimeter Array (ALMA), a world-wide project (64x12m-dishes operating from 84 to 720\,GHz, to be completed by 2011) will represent a jump of almost two orders of magnitude in sensitivity and angular resolution as compared to present millimeter/submillimeter interferometers, and will thus undoubtedly produce a major step in astrophysics. The main objectives will be the origins of galaxies, stars and planets. ALMA will be able to detect dust-enshrouded star-forming galaxies at redshifts $z$~$\ge$~10, both in the emission of  dust and spectral lines (CO and other species, including C$^+$). It will also explore in detail the physical and chemical processes of star and planet formation hidden away in dusty molecular clouds and protoplanetary disks. In addition, ALMA will allow similar enormous  gains in all other fields of mm and submm astronomy, including nearby galaxies, AGN, astrochemistry, circumstellar shells and the solar system.}  

\section{Introduction} 	
Millimeter (mm) and especially submillimeter (submm) astronomy is one of the youngest branches of astronomy. It has provided unique results about the cold interstellar medium and star formation, including proto--planetary disks, with the power of heterodyne technics, of high spatial resolution interferometry and of bolometer cameras. It has really opened the field of astrochemistry. More recently, it has proved its ability to detect dust and molecular emission at the highest known redshifts in Ultra-Luminous Infra-Red Galaxies (ULIRGs), tracing a major fraction of star formation in the Universe.

   	The outstanding capabilities of mm/submm radioastronomy for such purposes and for many others have led to consider major efforts to set up a new generation of facilities with order of magnitude increased capabilities. Advanced projects of large interferometers were thus first simultaneously studied in early 1990 in the United States (MMA), Europe (LSA) and Japan (LMSA). It was then decided to merge them in a unique worldwide project with much improved capabilities, the Atacama Large (sub)Millimeter Array (ALMA). The latter is currently a bi-partite joint project between North America (USA \& Canada) and Europe (ESO \& Spain) with the participation of Chile. It is foreseen that the Japan will become a partner in the ALMA project later on. The ALMA interferometer will be installed in an exceptional site for submm observations, at Chajnantor, Atacama, Chile, at 5010~m elevation, with baselines up to 14~km. It will include 64x12m--dishes, providing a very good coverage in the {\it uv} interferometric plane. It will operate in four frequency bands, covering four of the main atmospheric windows between 4\,mm and 0.4\,mm, and will be equipped with very broad band heterodyne receivers and correlators. Compared to existing facilities, its sensitivity and  angular resolution will be increased by almost two orders of magnitude. 

	This review summarizes the highlights of the ALMA project, (sub-)mm astronomy and the fields where major ALMA results are expected. Since lack of space precludes to detail all of them, it was chosen to give special emphasis to high redshift studies (see \cite{Gu01}, \cite{Wo01} and the ALMA web sites {\it http://www.eso.org/projects/alma/, http://www.alma.nrao.edu/}, for other aspects). 

\section{ALMA Project Summary}  
The Baseline ALMA Project has tasks and costs equally shared between the two main Partners North--America -- NSF/NRAO (+ NRC Canada) -- , and Europe -- ESO (+ Spain) --, while Chile provides the site. The key technical specifications are summarized in Table~1. The Phase I (development and design) has ended in 2002 and the project has now entered the construction phase (Phase II). The first antennas will arrive on the site in 2006. Early science will start at the end of 2007 with a limited number of antennas, and the array will be fully operational at the end of 2011.  

  	It is expected that Japan will join the project in 2004. In view of their participation, Japan has proposed to add to the baseline project a compact array of 12 x 7--m dishes, plus 4x12--m additional antennas dedicated to calibration, additional receivers and a second generation correlator.  

\section{The field of (sub-)mm astronomy: the cold universe, molecules and dust} 
 	Since ALMA studies will extend from $z$=0 to $z$$>$5, in the emission rest-frame the $\lambda$ range extends over almost two orders of magnitudes, from less than $\sim$70\,$\mu$m to 4\,mm. The physics of the photons detected by ALMA is thus the realm of molecular rotational transitions, plus cold dust and, more marginally, redshifted atomic and ionic fine structure lines.

  Millimeter \textbf{molecular lines} play the leading role for tracing the cold gas, its physical properties and the cosmic chemistry. More than 100 molecules have been identified by mm radioastronomy, mostly in the interstellar (IS) and circumstellar gas -- including rare isotopic varieties with D, $^{13}$C, $^{18}$O, etc. An updated list can be found on the web site {\it http://www.cv.nrao.edu/~awootten/allmols.html}. Formed mainly from H, C, O \& N atoms, they  can also include S, Si, metals, etc. They belong to two main classes:  expected common stable species made of up to $\sim$10 atoms; and a large number of exotic unstable species such as radicals, ions, long polyyne chains, small cycles, or isomers. Because of the difficulty of detecting  H$_2$, the CO molecule is the main tracer of the IS molecular gas.  

\begin{table} 
\caption[]{Key technical specifications of the ALMA Baseline Project} 
\begin{tabular}{ll}  \hline ARRAY					&		\\ 
Number of antennae		&	64 sx 12--m	\\ 
Total Collecting Area		&	7238 ~m$^2$	\\ 
Angular Resolution		& 0.2''~x~$\lambda$(mm)/baseline(km)$^\#$ \\ 
\hline 
CONFIGURATION			&		\\ 
Compact				&	150~m	\\ 
Highest Resolution		&	14~km	\\ 
\hline 
RECEIVERS			&		\\ 
Noise in 3\,mm \& 7\,mm bands&	20~--~50~K		\\ 
Noise in 2\,mm to 0.3\,mm bands&	6~h$\nu$/k SSB	\\ 
Windows &Band 3:  84--119 GHz		\\ 	
	&Band 6: 211--275 GHz	\\ 	
	&Band 7: 275--370 GHz 	\\ 	
	&Band 9: 602--720 GHz	\\ 
\hline 
BANDWIDTH 		&	8 GHz$^\#$, each polarisation	\\ 
\hline 
\hline 
SITE : Chajnantor (5010m)		&		\\ 
Median atmospheric transmission & $<$~0.05 at 225~GHz$^\#$	\\ 
Atmospheric phase calibration	& $<$~15 degrees rms$^\#$ \\ 
Amplitude calibration		&	1\%$^\#$	\\ 
\hline 
\end{tabular} \\ 
$^\#$Particularly outstanding specifications and qualities.\\ 
Explicit typical sensitivities for continuum and line detection of point sources can be found in \cite{Bl99} \& \cite{Gu01}, and for brightness temperature in  \cite{Gu01}; e.g., at ~230~GHz, in ~1~hour, ~~~\textbf{1-$\sigma$} is: $\sim$8~$\mu$Jy for point sources in continuum; $\sim$45~$\mu$Jy for point sources in line observations with a velocity resolution $\Delta$v~=~300 kms$^{-1}$ \cite{Bl99}; and $\sim$2.3x(10''/$\theta$)$^2$ $\mu$K for the brightness of extended sources \cite{Gu01}, where $\theta$ is the diameter of the synthesized beam. 
\label{} 
\end{table}

	These molecules allow various checks of physical and chemical IS processes. IS chemistry has been the main motor for the development of a new field, astrochemistry (see e.g. \cite{vD98}, \cite{Leq02}). As for other radio lines, including 21~cm HI, the heterodyne technics provide very precise velocity determination with many applications: disk rotation, virial mass determination of galaxies, inflows/outflows, jets, shocks, turbulence, kinematic distances in the Milky Way, etc.  

	\textbf{Fine structure lines} of abundant atoms and ions (C$^+$, 158\,$\mu$m; O{\small I}, 63\,$\mu$m and 146\,$\mu$m) are known to be the main coolents of the diffuse atomic IS gas. Only the important C{\small I} lines (370 and 609 $\mu$m) lie in the ALMA range at zero redshift. But at very high $z$, the lines of C{\small II}, O{\small I}, N{\small II} and O{\small II} also enter the ALMA range. In particular, the C$^+$ line -- prominent in spiral galaxies and moderate starbursts such as M~82 -- will be observable at $z$~$\ge$2, but it is relatively weak in strong starburst ULIRGs such as Arp 220 \cite{Lu98}. 

	For typical \textbf{dust} temperatures, T$_D$~$\sim$20--30~K, expected in  normal galaxies, the spectral energy distribution (SED)  peaks at $\approx$~100\,$\mu$m. However, T$_D$ is smaller, $\sim$10--20~K, for the cold gas, and higher in ULIRGs. A significant fraction of the SED always lies in the ALMA frequency range at any redshift. 

	With T~$>$~10$^3$--10$^4$~K, stellar and IS \textbf{plasmas} are not expected to be generally major  mm/submm targets; however, various studies are expected: stellar atmospheres, compact H{\small II} regions, AGN, Sunyaev--Zeldovich effect, etc. 

\section{Galaxy and star formation in the early Universe} 
    After the recent spectacular development of the exploration of the young Universe with HST and large ground--based  telescopes, all efforts are valuable to answer remaining basic questions about the nature of dark matter and dark energy, the physics of the first phases of Big Bang and the origins of structures in the Universe. The frontier of {\it detailed} knowledge of optical galaxies has now been pushed beyond $z$ $\sim$ 3 and is tackling $z$~$\sim$~6. This includes the main epoch of galaxy assembly, the peak of star formation and of quasar activity.  This frontier will soon be extended by the JWST well beyond $z$ = 10, to the dawn of the visible Universe when the first galaxies, stars and quasars were formed. A fundamental milestone in this quest is to understand the complex processes of galaxy formation and evolution, which are impossible to disentangle from the properties and history of star formation. It is now well proven that most of the UV radiation initially emitted by young stars at high $z$ is absorbed by dust in starburst galaxies and reemitted in the far-IR. The tentatives to correct for dust extinction the UV-visible radiation emerging from such high $z$ galaxies in order to infer the total star formation rate (e.g.~\cite{AS00}), have proven to lead to somewhat uncertain results, and a large fraction of the star formation energy is coming from ULIRGs hardly detectable even in the near-IR. A direct detection of the radiation emitted by dust and molecular gas at high $z$ is therefore mandatory to trace star formation in the early Universe. From $z$ $\sim$ 1 to $\ge$ 10, most of this emission is shifted into the submm/mm range where it is relatively easy to detect 

 \subsection{The power of (sub)mm studies for high $z$ starbursts} 

\textbf{~~~~a) Dust Detection}  
	In the optically thin Rayleigh--Jeans regime, the dust submm  spectrum is  extremely steep, varying as $\nu$$_o$$^{2+\beta}$, where $\beta$ $\approx$ 1.5--2 is the emissivity index. Therefore, at a fixed mm/submm detection frequency $\nu$, the emission luminosity at $\nu$$_0$ = (1+z)\,$\nu$, much increases with $z$. Then, despite the multiplication of the distance square factor by $\sim$1000, the detection sensitivity at a fixed $\lambda$ $\sim$ 1mm remains almost constant in the whole range of $z$ $\sim$0.5 -- 20 (see e.g. Fig. 4 of \cite{Co99}). This ``inverse K correction'' is a key aspect of the mm/submm window for the exploration of the Universe at very high $z$.  The  high $z$ mm/submm ULIRGs have a comoving density -- peaking at $z$ $\ge$ 2 -- several hundred times larger than at the present epoch. This leads to the unique and paradoxal result that mm/submm surveys are dominated by sources at $z$ $\ge$ 2. 

  In the last five years, a major breakthrough has been achieved with the bolometer camera SCUBA on the 15--m dish JCMT at Mauna Kea, in detecting high $z$ ULIRGs through wide-field surveys at 0.85\,mm (see e.g. references in \cite{Bl02}). Although, in the 1.2\,mm window, the flux of a typical ULIRG is lower by a factor $\sim$2--3, such a $\lambda$ range allows the use of larger telescopes and more ordinary sites. In such a way, the 117--channel MAMBO MPIfR bolometer camera  at the IRAM 30m telescope can complement SCUBA results by very wide 1.2\,mm surveys of strong sources \cite{Be00}.

  Altogether, a few hundreds of square arcminutes have thus been observed in SCUBA and MAMBO surveys, leading to the detection of more than 150  high $z$ sources. The best sensitivity achieved corresponds to far-IR luminosities of $\sim$10$^{12}$\,L$_\odot$. The corresponding source density, $\sim$1--2/arcmin$^2$, is not far from the confusion limit with the 14'' SCUBA beam. The detected sources account for an appreciable fraction of the (sub)mm background ($\sim$10\%), and they prove that such ULIRGs are major contributors to star formation at very high $z$. 

   However, such prominent submm sources remain difficult to identify at other $\lambda$, and only a part of them have been  detected in deep near-IR observations allowing redshift determination (\cite{Bl02}, Ivison et al. in preparation). Nevertheless, from the ratio of the (sub)mm and radio fluxes, it has been shown that most sources of SCUBA--MAMBO surveys have $z$ between 1 and 4 \cite{CY99}. In fact, there is yet practically no such ULIRG well identified at $z$$>$4, except those with (sub)mm detections from pointed observations toward bright optical AGN, i.e. more than 50 sources at $z$$>$4, including a few radiogalaxies and mostly bright QSOs (see references in \cite{Om03}). Although part of this emission may be direct reprocessing by dust of the AGN radiation (see e.g. \cite{An99}), there is evidence in many cases, from radio and CO emission, that an appreciable part comes from  dust heating by starburst (as in ULIRGs without AGN), with far-IR luminosities in the range 10$^{12}$--10$^ {13}$\,L$_\odot$. 

\textbf{b) CO Detection}
  	A similar ``negative K--correction'' also applies for high $z$ CO line emission, the best tracer of molecular gas and thus of starbursts and of dynamical masses. If one considers the detection in a given spectral range, e.g. the 3 mm atmospheric window, increasing $z$ means to address a higher J transition in the rotation ``ladder'' (e.g. J=5--4 for $z$$\sim$4--5). Since the line strength much increases with J, there is again a very efficient compensation for  the distance square attenuation of the line intensity (see e.g. Fig. 4 of \cite{Co99}). 

   The detection of mm lines from high $z$ ULIRGs is presently more difficult than for dust because of the advantage of the very broad band of bolometers, opacity effects \cite{Co99} and of the need to have accurate $z$ value for line searches. In fact, CO has been currently detected only in $\sim$\,20 sources at $z$~$>$~1 (see e.g. Table 1 of \cite{Cox02}). For various reasons -- technical, atmospheric, line excitation -- the best sensitivity is achieved in the 3\,mm range. Interferometers have proven to be much better than single dishes for detecting very weak extra-galactic lines, because their baseline is much less affected by instrumental ripples which can mimic fake line detections, and the spatial concentration of the line emission at the source position is a strong argument to distinguish them from spurious cases.  

   Despite the importance of such results, many questions remain to be addressed by ALMA about the properties of high $z$ ULIRGs and their contribution to the history of star formation in the Universe, e.g.: how are they spatially distributed ? Can we find evidence of protoclusters ? What are the properties of those harbouring AGN, and especially X-ray detected obscured AGN ? What is the nature of weaker submm sources and their relation with the other main contributors to high $z$ star formation, Lyman--break galaxies ? What are their dynamical mass and their structure -- circumstellar disk, bulge, galactic disk, galaxy merging, etc. -- , tracing the history of their formation ?   

\subsection{ALMA capabilities at high $z$}  
\textbf{~~~~a) Sensitivity.} With a sensitivity for dust detection at 850\,$\mu$m $\sim$\,50 times better than SCUBA and a survey speed close to 1000 times faster, ALMA will allow a big leap forward in surveying high $z$ starburst galaxies, without being limited by source confusion (see e.g. \cite{Bl99}, \cite{Bl02}, \cite{Co00}, \cite{La03}, \cite{Wo01}). For instance, a 4'\,x\,4' field could be observed to a sensitivity of 0.1 mJy (20 times lower than current SCUBA field surveys) in two weeks, yielding the detection of $\sim$100-300 galaxies, together with CO detection in many of them. This will permit the detection of moderate starbursts ($\sim$10$^{11}$~L$_\odot$), up to $z$~$\approx$~10-20, if any, i.e. {\it the first starbursts in the Universe}.  

	Shallower surveys will be also performed in 10-100 times larger areas, but not exceeding about one square degree. Special fields will be also addressed with the unique ALMA sensitivity, in particular lensed fields by clusters of galaxies, and fields with hints of high $z$ proto-clusters. ALMA sensitivity will determine the far-IR luminosity and the molecular gas content of many particular, exceptional, often gravitationally lensed, high $z$ galaxies discovered e.g. by HERSCHEL, JWST and surveys at various $\lambda$ (IR, X-ray, radio, optical, etc.). Altogether, ALMA will certainly detect or observe several thousand high $z$ galaxies, most of them too weak to be detected by other mm/submm facilities.  

\textbf{b) Very large bandwith.} The 2x8~GHz bandwidth will cover the whole 3\,mm window in only two different frequency tunings. This will allow in particular blind searches for CO in sources with completely unknown $z$, and thus {\it redshift determination} in many dust detected ULIRGs where $z$ measurement at other $\lambda$ is extremely difficult. For strong emitters, multi-line studies of various species will be also much faster, allowing chemical studies at high $z$.  Similarly, the combination of high sensitivity, very small beam and moreover very large bandwidth will make ALMA a unique facility for line absorption studies along the line of sight of strong (sub)mm continuum sources \cite{CW98}, in particular when the absorption redshifts are unknown.  

\textbf{c) High submm frequencies.} The determination of the whole submm SED will allow one to directly infer fundamental properties of the detected high $z$ ULIRGs: {\it submm photometric redshifts}, dust temperatures, {\it far-IR luminosities}, etc. As discussed in Section 3, an important goal of ALMA will be the comprehensive study of the {\it C$^+$ fine-structure line} in high $z$ galaxies, especially with moderate starbursts.  

\textbf{d) High resolution imaging.} The exquisite angular resolution of ALMA, better than 0.1'' ($<$$\sim$~1\,kpc at $z$~$>$~1), will fully resolve high $z$ galaxies, including their bulges, with their velocity structure in CO lines. The determination of {\it dynamical masses} from CO lines will be a major outcome to constrain galaxy evolution. Special cases of high lensing magnifications will allow addressing more details and weaker sources.  

\textbf{e) Polarisation.} In addition to (sub)mm polarisation studies of high $z$ radiosources, the polarisation capability of ALMA could be useful to try to trace magnetic fields in high $z$ ULIRGs.   

 \section{Proto-stars and planet formation}  
	The large increase in sensitivity and angular resolution of ALMA will bring major progress in all traditional fields of mm astronomy. The broad and complex problem of star formation is the first one, with its various aspects from large scale to formation of planetary systems. The angular resolution of ALMA should be decisive for breakthroughs in the latter. It is impossible to review here all the others. We only list below a few of them, the  most significant ones.	  

\subsection{The interstellar medium and star formation}  

\textbf{~~~~a) Fractal structure of IS clouds} (\cite{He98} and references therein). The high resolution of ALMA will trace the threshold of the structure hierarchy, down to 10 AU.  

\textbf{b) Millimeter absorption spectroscopy} of IS clouds  (\cite{Li00} and references therein). The much higher of ALMA sensitivity will increase the available (extragalactic) background radiosources by two orders of magnitude. Many developments will be possible in studies of translucent molecular clouds (chemistry, time variations, etc.), with extensions to nearby galaxies.  

\textbf{c) Dense prestellar cores.} The initial conditions of gravitational collapse will be determined in nearby star forming regions, down to Jupiter mass. Much information will be gathered about larger prestellar cores up to the Galactic Center.

\textbf{d) Molecular clouds in nearby galaxies.} The observation of many  more details will allow systematic comparison with respect to star formation in the Milky Way, in various environments (metallicity, age, merging, galaxy clusters, etc.)  	

 \subsection{Young stellar objects (YSOs)}  
\textbf{~~~~a) Structure of protostellar envelopes and disks} (``Class 0''). One can expect systematic direct studies of the gas infall, with disentangling of infall, outflow and rotation. The capability to observe weak submm lines and dust emission will be fundamental to study the hotter inner regions.  

\textbf{b) Molecular outflows} will be mapped with exquisite details about shocks, magnetic field, structure and chemical evolution, etc.  

\textbf{c) Fragmentation} and multiple star formation will be especially addressed in nearby star forming regions, in relation with the current progress about the IMF and the census of very low mass stars and substellar objects.   

\subsection{Proto-planetary disks}  	
ALMA should bring decisive progress in understanding the processes of formation and evolution at various stages of evolution of (proto-)planetary disks, preceding and accompanying the formation of the various planets and their subsequent evolution \cite{Gu01}.  

\textbf{a) Massive disks around pre-main sequence stars}. Such disks seem to accompany the formation of most low mass stars (T Tauri and Herbig Ae stars) and even sub-stellar objects down to $\sim$0.01 M$_\odot$. ALMA will detect all the dust in such nearby disks, as well as the gas mostly through CO lines. In addition to the extended thin outer layers, the ALMA sensitivity (down to a few Earth mass of dust and gas) and resolution will apply to the inner more active part of the disk, $\sim$10~AU. Binary systems will be particularly interesting with the possibility of circumbinary disks, tidal gaps created by the secondary stars, etc. {\it Tidal gaps} created in planetary disks by external planets will be specially exciting targets as direct signposts of planets.  

\textbf{b) Dispersal of gas and dust in disks.} The ALMA sensitivity will allow following the whole phase of evolution and the dispersion of disks from a few million to several ten million years. The presence and the distribution of dust and CO will thus be checked in a large sample of disks around various stars.  

\textbf{c) Debris disks and zodiacal dust systems.} The few disks detected by IRAS around zero age main sequence stars ($\beta$ Pic, Vega, etc.) are clearly identified as debris disks from the destruction of larger bodies, with some evidence of planet presence \cite{Lec98}. Their dust emission has been detected by SCUBA, close to its sensitivity limit. The ALMA sensitivity is required for detailed mapping of dust, mm detection of CO, and systematic studies of many more disks to be detected by SIRTF and ASTRO-F.  

\textbf{d) Chemistry in protoplanetary disks.} The chemistry in such complex and rapidly evolving objects look specially appealing. One expect strong chemical gradients, with many specific processes: gas depletion by accretion into grain mantles, photo-processing of mantles, grain desorption possibly of complex species; ion-molecule reactions; photochemistry; X-ray irradiation of grains and gas, etc.    


\subsection{Galaxies and AGN}  
 
\textbf{~~~~a) Nearby galaxies.} The high resolution of ALMA will allow  mapping all details of galaxy structure (arms, bar, ring, etc.) in CO lines up to the distance of the Virgo Cluster and beyond. The detection of individual molecular clouds in several lines, in particular CO isotopes, will allow comparative detailed studies of the molecular gas and star formation in large samples of various types of galaxies, e.g.: galaxies in clusters and various groups; cold gas in elliptical galaxies tracing merger history; starbursts and peculiar cases of mergers; dwarf galaxies and molecular clouds outside of galaxies; etc.  

\textbf{b) Local Group and Magellanic Clouds.} Very small structures, smaller than one parsec, will be detectable in the whole Local Group, and as small as 10$^{16}$~cm in the Magellanic Clouds. Standard studies currently performed in the Milky Way will be extended  to the Magellanic Clouds allowing comparative studies: structure and chemistry of molecular clouds, star formation, diffuse gas, circumstellar SiO masers, etc. 

\textbf{c) AGN} ALMA could extend present studies of the circumnuclear molecular gas to all the known Seyfert galaxies, and give high resolution maps of the innermost parts of the circumnuclear
disks. It will be able to map not only the gas but also the dust obscuring the galaxy nuclei. This will give us greatly improved
possibilities of studying the connection between star formation and massive black holes in nearby AGN.

\textbf{d) Ionisation fluctuations, Sunyaev-Zeldovich effect}  
The mm range is ideal for detecting ionized cosmic structures by their imprint on the cosmic background, with a sensitivity independent of distance (Sunyaev-Zeldovich [SZ] effect). However, the best cases, galaxy clusters have dimensions, $>\sim$1', much larger than ALMA synthesized beam ($<$ a few arcsec). Therefore, ALMA will not be directly useful for the SZ detection of clusters; but it will be powerful in studying smaller ($\sim$3--10'') ionized structures such as: cluster substructures (mergers, ``fronts'', etc.) revealed by X-ray studies (see e.g. Forman et al. in this volume), which should be particularly pronounced at high $z$; or even isolated structures, around (proto-) goups or galaxies, if the sensitivity and confusion allow.

\subsection{Stars and their evolution}  
\textbf{~~~~a) Millimeter continuum emission from stars.} Mainly thermal emission may be detected from photospheres, chromospheres and winds of various stars, with extension of the studies to much larger distances and many applications: wind acceleration; B[e] shells; stellar activity; extended atmospheres of giants and supergiants; symbiotic stars; accretion shocks and non-thermal processes; etc.

\textbf{b) Circumstellar envelopes of AGB stars.} The sensitivity and angular resolution of ALMA will open new opportunities for measuring mass-loss rates and mass-return to the IS medium and tracing time-dependent chemistry in peculiar (carbon-rich) conditions. Main challenges for ALMA will include: the chemistry of dust formation by high-resolution submm observations; extending CO mass-loss determinations through Galactic Center distances and marginally to the Magellanic Clouds; high-resolution mapping of a large sample of (peculiar) shells, for studying CO and dust spatial distribution and localised chemistry processes; tracing mass-loss variations on time scales from 100 to 10$^5$~yr; mass-loss determinations in binary systems; evidence of planet--envelope interactions; systematic studies of SiO masers in the Magellanic Clouds; etc.   

\textbf{c) Post-AGB sources.} Circumstellar envelopes of \textbf{proto--planetaty nebulae} share many properties of AGB envelopes, with rich peculiarities: hollow shells, super-winds and fast winds, bipolar structure, peculiar chemistry, etc. ALMA will be unique for studying these fast-evolving, rare, peculiar objects, in particular for high-resolution observations of their structure and dynamics, in most of the Galaxy. Similarly, in the next stage of evolution, \textbf{planetary nebulae}, ALMA will be unique for detection and high-resolution studies of their peculiar, dense and warm molecular gas.  

\textbf{d) Astrometry.} The high resolution of ALMA will allow proper motion studies of these various classes of sources up to Galactic Center distances. Such capabilities will be particularly precious in obscured regions, and, e.g., systematic studies of SiO masers will address the structure of the inner Milky Way. (Sub)mm VLBI including ALMA could extend  astrometry capabilities for strong sources.  

\subsection{The Sun and the Solar System}  
\textbf{~~~~a) Comets and asteroids.} ALMA will allow obvious extensions of present studies of the physics and chemistry of cometary atmospheres: high resolution, larger distances, detection of larger molecules, simultaneous observations of many rare species and isotope varieties, time variations, velocity determination, etc. One expects much deeper insight in the composition of cometary ices and the origin of comets. ALMA will also address surface properties of small bodies: asteroids, comet nuclei, Trans--Neptunian Objects.  

\textbf{b) Planetary atmospheres.} ALMA will have the capability of 3D instantaneous mapping -- with angular resolution down to 100~km and a unique coverage in the {\it uv} interferometric plane -- of the dynamical structure, the chemical composition, the temperature, the winds, etc. Most important studies should be: in Mars and Venus: winds, jets, clouds, water cycle in Mars, etc; in giant planets: chemistry, winds, D/H ratio, PH$_3$, etc.; Titan chemistry; tenuous atmospheres of Io, Pluto, Titon, etc.

\textbf{c) The Sun.} The high angular resolution of ALMA may be useful for mapping solar flares and determining the structure and dynamics of the chromosphere, mostly in the continuum, but also in recombination lines.   

 \subsection{Cosmic chemistry}  As mentioned  above, ALMA will bring significant advances in the chemistry of practically every molecular medium in the Universe, from high $z$ galaxies to cometary atmospheres. Because of its high angular resolution, its impact will be particularly high for localised chemical processes such as: circumstellar disks, protostars, shocks, dust formation in circumstellar envelopes, AGN tori, cometary atmospheres, etc. Dust chemistry, probably the most difficult branch of cosmic chemistry, will thus greatly benefit, both for grain formation and accretion and for desorption products. The very broad bandwidth of ALMA will permit very fast complete surveys of molecular lines in a large number of sources, yielding comprehensive views of the chemistry and allowing comparative studies. The opening of submm windows will be of special interest for hydrids, including deuterated species, and for the chemistry of hot media.

   A central question is whether ALMA will produce substantial progress in the general IS chemistry, by reaching a new step in the detection of prebiotic species. After the fast discoveries of new IS molecules in the seventies and eighties, their rythm has slown down in the last years. Even glycine has not yet been firmly detected despite many efforts for searching it (see e.g. \cite{Sn97}). It is clear that, in the best studied sources, one is already not far from the confusion limit by weak lines of various species which occupy most of the spectral range. However, ALMA will be the best tool for further systematic efforts justified by the fundamental importance of the issue. There seems to be a very good chance that well focused, organised programs will produce the detection of at least a few prebiotics molecules, in particular amino--acids, such as those currently identified in meteorites (see references in \cite{Sh02}). It is clear that it will be necessary to get complete, homogeneous spectra with the best possible signal-to-noise ratios, allowing sophisticated data processing to extract characteristic spectra of given species. The sources should be best chosen for synthesizing such species and for minimizing line confusion. Since the synthesis of such complex molecules by irradiation on grains is a possible process, regions of grain desorption could be specially addressed.  

\section{Conclusion}
First world-wide project of this level in astronomy, ALMA will play a role similar to HST in the optical by opening a completely new range of observational possibilities. There will be no equivalent until ALMA is operational, and it will cover nearly all major topics in astronomy from comets to cosmology, from astrometry to exo--biology, with a special impact in understanding the origins of galaxies, stars and planets, and highlights such as : extremely deep unconfused surveys for normal galaxies at z~$>$~5, detecting galaxies up to several hundred times fainter than those detected using SCUBA; resolving luminous and gravitationally lensed galaxies, allowing their internal dynamics to be traced in great detail; revolutionizing the study of formation of stars and protoplanetary disks by providing a combination of angular resolution and sensitivity that far exceeds that of present instruments, placing crucial constraints on planet formation theories. The dramatically enhanced sensitivity and resolving power of ALMA will allow great advances in all fields, perfectly complementing the optical and infrared performance of the JWST.

\medskip
{\it Acknowledgements.} I  thank Pierre Cox  for his  careful help in preparing this manuscript, and Martin Giard for helpful discussions.

\end{document}